\documentclass{ws-p8-50x6-00}

\begin{document}

\newcommand\vev[1]{\langle{#1}\rangle}

\def\la{\mathrel{\mathpalette\fun <}}
\def\ga{\mathrelbe {\mathpalette\fun >}}
\def\fun#1#2{\lower3.6pt\vbox{\baselineskip0pt\lineskip.9pt
        \ialign{$\mathsurround=0pt#1\hfill##\hfil$\crcr#2\crcr\sim\crcr}}}

\renewcommand\({\left(}
\renewcommand\){\right)}
\renewcommand\[{\left[}
\renewcommand\]{\right]}

\newcommand\del{{\mbox {\boldmath $\nabla$}}}

\newcommand\eq[1]{Eq.~(\ref{#1})}
\newcommand\eqs[2]{Eqs.~(\ref{#1}) and (\ref{#2})}
\newcommand\eqss[3]{Eqs.~(\ref{#1}), (\ref{#2}) and (\ref{#3})}
\newcommand\eqsss[4]{Eqs.~(\ref{#1}), (\ref{#2}), (\ref{#3})
and (\ref{#4})}
\newcommand\eqssss[5]{Eqs.~(\ref{#1}), (\ref{#2}), (\ref{#3}),
(\ref{#4}) and (\ref{#5})}
\newcommand\eqst[2]{Eqs.~(\ref{#1})--(\ref{#2})}

\newcommand\pa{\partial}
\newcommand\pdif[2]{\frac{\pa #1}{\pa #2}}

%units
\newcommand\yr{\,\mbox{yr}}
\newcommand\sunit{\,\mbox{s}}
\newcommand\munit{\,\mbox{m}}
\newcommand\wunit{\,\mbox{W}}
\newcommand\Kunit{\,\mbox{K}}
\newcommand\muK{\,\mu\mbox{K}}

\newcommand\metres{\,\mbox{meters}}
\newcommand\mm{\,\mbox{mm}}
\newcommand\cm{\,\mbox{cm}}
\newcommand\km{\,\mbox{km}}
\newcommand\kg{\,\mbox{kg}}
\newcommand\TeV{\,\mbox{TeV}}
\newcommand\GeV{\,\mbox{GeV}}
\newcommand\MeV{\,\mbox{MeV}}
\newcommand\keV{\,\mbox{keV}}
\newcommand\eV{\,\mbox{eV}}
\newcommand\Mpc{\,\mbox{Mpc}}

%astronomical
\newcommand\msun{M_\odot}
\newcommand\mpl{M_{\rm P}}
\newcommand\MPl{M_{\rm P}}
\newcommand\Mpl{M_{\rm P}}
\newcommand\mpltil{\widetilde M_{\rm P}}
\newcommand\mf{M_{\rm f}}
\newcommand\mc{M_{\rm c}}
\newcommand\mgut{M_{\rm GUT}}
\newcommand\mstr{M_{\rm str}}
\newcommand\mpsis{|m_\chi^2|}
\newcommand\etapsi{\eta_\chi}
\newcommand\luv{\Lambda_{\rm UV}}
\newcommand\lf{\Lambda_{\rm f}}

\newcommand\lsim{\mathrel{\rlap{\lower4pt\hbox{\hskip1pt$\sim$}}
    \raise1pt\hbox{$<$}}}
\newcommand\gsim{\mathrel{\rlap{\lower4pt\hbox{\hskip1pt$\sim$}}
    \raise1pt\hbox{$>$}}}

\newcommand\diff{\mbox d}

\def\dbibitem#1{\bibitem{#1}\hspace{1cm}#1\hspace{1cm}}
\newcommand{\dlabel}[1]{\label{#1} \ \ \ \ \ \ \ \ #1\ \ \ \ \ \ \ \ }
\def\dcite#1{[#1]}

\def\calm{{\cal M}}
\def\calp{{\cal P}}
\def\calr{{\cal R}}
\newcommand\calpr{\calp_\calr}                        

\newcommand\bfa{{\bf a}}
\newcommand\bfb{{\bf b}}
\newcommand\bfc{{\bf c}}
\newcommand\bfd{{\bf d}}
\newcommand\bfe{{\bf e}}
\newcommand\bff{{\bf f}}
\newcommand\bfg{{\bf g}}
\newcommand\bfh{{\bf h}}
\newcommand\bfi{{\bf i}}
\newcommand\bfj{{\bf j}}
\newcommand\bfk{{\bf k}}
\newcommand\bfl{{\bf l}}
\newcommand\bfm{{\bf m}}
\newcommand\bfn{{\bf n}}
\newcommand\bfo{{\bf o}}
\newcommand\bfp{{\bf p}}
\newcommand\bfq{{\bf q}}
\newcommand\bfr{{\bf r}}
\newcommand\bfs{{\bf s}}
\newcommand\bft{{\bf t}}
\newcommand\bfu{{\bf u}}
\newcommand\bfv{{\bf v}}
\newcommand\bfw{{\bf w}}
\newcommand\bfx{{\bf x}}
\newcommand\bfy{{\bf y}}
\newcommand\bfz{{\bf z}}

\newcommand\sub[1]{_{\rm #1}}
\newcommand\su[1]{^{\rm #1}}

\newcommand\supk{^{(K) }}
\newcommand\supf{^{(f) }}
\newcommand\supw{^{(W) }}
\newcommand\Tr{{\rm Tr}\,}

\newcommand\msinf{M\sub{inf}}
\newcommand\phicob{\phi\sub{COBE}}
\newcommand\mgrav{m_{3/2}(\phi)}
\newcommand\mgravsq{m_{3/2}^2(\phi)}
\newcommand\mgravcu{m_{3/2}^3(\phi)}
\newcommand\mgravvac{m_{3/2}}

\newcommand\pone{\dot\phi_1}
\newcommand\ptwo{\dot\phi_2}
\newcommand\ponesq{\dot\phi_1^2}
\newcommand\ptwosq{\dot\phi_2^2}
\newcommand\meff{m\sub{eff}}

\newcommand\cpeak{\sqrt{\tilde C_{\rm peak}}}
\newcommand\cpeako{\sqrt{\tilde C_{\rm peak}^{(0)}}}
\newcommand\omb{\Omega\sub b}
\newcommand\ncobe{N\sub{COBE}}

\newcommand\sigtil{\widetilde\sigma_8}
\newcommand\gamtil{\widetilde\Gamma}

%%%%%%%%%%%%%%%%%%%%%%%%%%%%%%%%%%%%%%%%%%%%%%%%%%%%%%%%%%%%%%%%%%%%%%%%%%

\title{ Observational constraints on models of inflation from
  the  density perturbation and  gravitino production\footnote
{Updated version of a talk at COSMO2K, to appear in the proceedings}}
\author{ David H. Lyth} 
\address{Department of Physics\\Lancaster University, Lancaster LA1 4YB, U.K.}

\maketitle

\abstracts{Present data require a spectral index $n\gsim 0.95$
at  something like 1-$\sigma$ level. If this lower bound survives
it will constrain `new' and `modular' inflation models, while  raising it
to $1.00$ would rule out all of these models plus  many others.
 After inflation, gravitinos are created by the oscillating field until  
the `intermediate' epoch when the Hubble parameter falls below the 
gravitino mass, or reheating, whichever is earlier. In a wide range 
of parameter space, these gravitinos are more abundant than those from 
thermal collisions, leading to stronger cosmological constraints.}

\section{Introduction}
Inflation is supposed to do two quite different jobs \cite{treview,book}.
Starting at the Planck scale, inflation should protect  our
patch of the universe against 
collapse, and against invasion by the presumably hostile region
around it. Then, much later, when the rate of expansion
 is at least five orders of magnitude below the Planck
scale, inflation is supposed to generate the specific 
 initial conditions, that are required for the subsequent
Hot Big Bang if it is to produce the observed Universe.
The second job is done during the last 70 (or fewer) $e$-folds of inflation.
Only that era is directly accessible to observation, and it is the focus
of this article.
The initial conditions  include the following
\begin{itemize} 
\item   An extremely homogeneous and isotropic Universe
\item A spatially flat Universe 
\item  A clean Universe:
 no relics which would spoil nucleosynthesis, overclose the Universe,
or otherwise contradict observation.
  \item    A primordial curvature  perturbation, whose spectrum
is rather flat on cosmological scales.
\end{itemize}
Inflation sets the first and  second  conditions in a completely
straightforward fashion. The same is true of the
fourth condition, provided that the inflation is of the slow-roll
variety. The  third condition, though, might be problematic
because the slow-roll inflation which generates the curvature perturbation
may also generate light relics with gravitational-strength interactions,
such as   moduli with spin 0, and the gravitino with spin $3/2$.
To get rid of these one might need 
 a separate bout of late inflation, 
lasting only a few $e$-folds. According to present thinking, the
late inflation would not be of the slow-roll variety, but rather
what is called \cite{thermal} thermal inflation.

This article is  in two parts. In the first part, I focus on
the primordial density perturbation, and in particular on
the spectral index which specifies the scale-dependence of its
spectrum. 
 The spectral index  
 is a potentially very powerful
 discriminator between different inflation models. Already, 
the constraint on the spectral index from a best fit to
relevant data is on the verge of ruling out hitherto popular
models. 

In the second part of the article, I discuss the
production of gravitinos after slow-roll inflation.
On the basis of the rather complete formalism recently presented by Kallosh
et al.~\cite{kklv2}, it has been shown recently  \cite{00grav} that
 the  conjecture \cite{latetime}
of late-time gravitino creation
is likely to be correct:  gravitinos are created at the `intermediate
epoch' when the Hubble parameter falls below the gravitino mass, or
at reheating, whichever is earlier.
 This bout of late-time creation usually swamps the bout
of creation just after inflation \cite{kklv,gtr}, leading to 
stronger cosmological constraints.

\section{Inflation and the spectral index of the
 primordial curvature perturbation}

\begin{table}[t]
\caption{A brief history of the Universe.
}
\begin{center}
\footnotesize
\begin{tabular}{|l|l|}
\hline
 (energy density)${}^\frac14$ & \\
\hline
$10^{18}\GeV$? & Inflation begins \\
$10^{13}\GeV$?? &  Primordial curvature perturbation freezes  \\
&  Inflation ends soon afterwards \\
 & We {\sl don't know} what happens next, until \dots \\
$ 1\MeV$ & Nucleosynthesis \\
$1\keV$ &  Primordial curvature perturbation unfreezes \\
&   Matter becomes clumpy \\
& Radiation becomes anisotropic  \\
$10^{-3}\eV$ & Present epoch \\
\hline
\end{tabular}
\end{center}
\label{table1}
\end{table}

Let us begin by recalling 
the history of the Universe, as summarized in Table \ref{table1}.
The curvature perturbation is generated when cosmological scales
leave the horizon during inflation. Until these scales re-enter the
horizon, long after inflation, it is time-independent (frozen in);
this is the object that I am calling the {\em primordial}
 curvature perturbation. 
The freezing-in of the curvature perturbation on super-horizon scales
is a direct consequence of the lack of causal interactions on such scales,
under the sole assumption of energy conservation \cite{llmw}, and independently
of whether Einstein gravity is valid. This is 
 extremely fortunate,
 since we know essentially nothing
 the Universe while cosmological scales are outside the horizon.

The 
 spatial Fourier components of the primordial curvature perturbation
are uncorrelated (Gaussian perturbation),
which means that its stochastic properties are completely determined
by its {\em  spectrum}  $\calp_\calr(k)$, 
 defined essentially as the mean-square value of the 
spatial Fourier component with comoving wavenumber $k$.
The  {\em spectral index} 
\be
n(k)\equiv 1 +  \frac{\diff \log \calp_\calr}{\diff \log k}
\nonumber
\ee
defines the shape of the spectrum.

A {\em special case}, predicted by  most inflation models,
is that of a practically scale-invariant $n$, giving
 $\calp_\calr \propto k^{(n-1)}$. 
 The {\em most special case}, predicted only by rather special  models
of inflation, is that of a spectral index practically indistinguishable
from $1$, giving a practically scale-invariant $\calp_\calr$.

By the time that cosmological scales re-enter
the  horizon, long  after nucleosynthesis,
 we know the content of the Universe; there are photons, three types
of neutrino with (probably) negligible mass, the baryon-photon fluid,
  the 
(non-baryonic) dark matter, and the cosmological constant.
The 
 primordial curvature perturbation is associated with
  perturbations in the
densities of each of these  components, which all
vanish on a common 
spatial slicing (an adiabatic density perturbation). It is also associated
with anisotropies in the momentum distributions. Using well-understood
coupled equations, encapsulated say in the CMBfast package,
the perturbations and anisotropies can be evolved
forward to the present time, if we have a well-defined cosmological model.
Here we will make the simplest assumption, namely the
 $\Lambda$CDM cosmology; the  Universe is spatially flat, and the
 non-baryonic cold dark matter is cold (CDM). 
Flatness is the naive prediction of 
inflation, and there is no  definite evidence against CDM.

I would like to report the result of a recent fit \cite{cl00}
 of the parameters
of the $\Lambda$CDM model. 
The data set consisted of the following.
\begin{itemize}
\item The normalization $(2/5)\calp_\calr^{1/2}=1.94\times 10^{-5}$ from
COBE data on the cmb anisotropy.
\item Boomerang and Maxima data at the first and second peaks of the
cmb anisotropy.
\item Hubble parameter $h=0.65\pm0.075$, 
total density $\Omega_0=0.35\pm0.075$, 
baryon density $\Omega\sub B h^2=0.019\pm 0.002$.
\item Slope of galaxy correlation functions $\gamtil=0.23\pm 0.035$
\item RMS matter density contrast  $\sigtil=0.56\pm0.059$ in sphere of radius
 $8h^{-1}\Mpc$
\end{itemize}
The epoch of reionization  was {\em calculated},
 assuming that a  fraction
 $f\gsim 10^{-4}$ has  collapsed.

The result  (for $f\simeq 10^{-2}$) is
\be
n=0.99\pm0.05
\label{nresult}
\ee
 This is higher than that of Kinney et al. \cite{mkr}
($n=0.93\pm 0.05$)
 and of Tegmark et al. \cite{teg00}
($n=0.92\pm0.04$).
Probably, this is because the former 
 do not include $\sigtil$ or $\gamtil$, while the latter
do not include $\sigtil$, and have also a  lower $\gamtil$.
Also, both have  reionization redshift $z\sub R\simeq 0$.
We shall see that the tighter lower bound on $n$ implied by our analysis
is  significant, in the context of 
some models of inflation. (These are the only two analyses so far which
include most of the relevant data, including the crucial nucleosynthesis
constraint. A recent analysis \cite{dick00}
omitting the latter gives $n=1.03\pm0.08$.)

\section{Comparison with models of slow-roll inflation}

The near scale-independence of the primordial curvature perturbation
presumably requires slow-roll inflation, in which the potential
$V$
satisfies flatness conditions $\mpl|V'/V|\ll 1$ and
$\mpl^2|V''/V|\ll 1$. (We do not consider the possibility 
of a break
in the spectrum, associated with temporary failure of slow-roll; see
for instance \cite{bgss}.)
Assuming 
a single-component inflaton and Einstein gravity, the 
prediction depends mainly  on the inflaton potential $V(\phi)$,
and the 
 number of $e$-folds $N\sub{COBE}$ of slow-roll inflation after
the scales explored by COBE  leave the horizon.
(It  depends also on  the 
 inflaton field value $\phi\sub{end}$ when slow-roll ends, but 
in an interesting class of models this dependence is very weak.)
The prediction of slow-roll inflation is
\bea
\frac4{25}\calp_\calr(k) 
&=& \frac1{75\pi^2\mpl^2} \frac{V^3}{V'^2} \nonumber\\
 \frac{n(k)-1}{2} & =&  \mpl^2 \frac{V''}{V}
 - 3\mpl^2 \( \frac{V'}{V} \)^2
\nonumber 
\eea
 The right hand side is to be evaluated at the epoch of
 horizon exit $k=aH$, given by
\bea
N(k)&\equiv& \ln(k\sub{end}/k)
 = \mpl^2 \int^\phi_{\phi\sub{end}} \frac{V}{V'} \diff \phi \nonumber\\
N(k\sub{COBE}) &=& 60 - \ln\frac{10^{16}\GeV}{V^\frac14} -
\frac13 \ln\frac{V^{1/4}}{T\sub R}- N_0 \nonumber
\eea

The number
$N_0$ (non-negative in any reasonable cosmology) parameterizes our ignorance
about the history of the Universe between the end of slow-roll inflation
and  nucleosynthesis. It is zero
 in standard cosmology,  but one bout of  thermal inflation \cite{thermal}
could generate $N_0\sim 10$, and two or more bouts are quite 
feasible.

\subsection{Models of slow-roll inflation}

The easiest way of satisfying the flatness conditions is to have
field values $\phi\gg\mpl$; then the flatness conditions are satisfied
by  $V=V_0f(\phi/\mpl)$, where $f$ is any function whose value and derivatives
are of order 1. In a non-hybrid model, $f$ should becomes steep so that inflation ends, but that still leaves a lot of freedom. The simplest choice is a
monomial $V\propto \phi^2$ or $\phi^4$ (usually called chaotic inflation), 
but in the large-field
regime monomials have no special significance. The reason is that all of the
 coefficients in  a power series for $V$ are expected to have coefficients
of order 1 in Planck units.

In principle,  string theory presumably 
determines these  coefficients.
 This idea  does
yield some proposals for the potential of special fields such as 
moduli \cite{treview,bd}
 or fields corresponding to the distance between D-branes \cite{gia}.
In the former case one might have inflation with the potential in the
last row of Table 2. With this possible exception, it seems that
if Nature has chosen to inflate at large
field values, there is at present no theoretical guidance
about the form of the potential. 

For this reason, models of inflation
based on current theoretical ideas \cite{treview} should
invoke $\phi\lsim\mpl$ and preferably
$\phi\ll\mpl$. Then it is justified to focus on 
 the  renormalizable terms of the potential (quartic and lower). 
(One or two  non-renormalizable terms might be invoked for 
special purposes.)
By inflating along a flat direction of global supersymmetry,
the flatness conditions are marginally satisfied. 
To have them well-satisfied, without  fine-tuning, one can invoke
an approximate
 global symmetry $\phi\to\phi+$const (shift symmetry).
With any such symmetry, the
 potential is completely flat  in the limit
of exact symmetry, and (provided that the potential does not vanish in this 
limit) the approximate flatness required for inflation
can be ascribed to the approximate symmetry \cite{treview,ewan}. 

The difference between the large- and small field cases is
 neatly illustrated by a proposal \cite{kawasaki}
 reported at this meeting. It seeks to justify the 
potential $V\propto \phi^2$ at $\phi\gg\mpl$ by invoking
a shift symmetry, but to achieve this a particular symmetry-breaking
term is invoked. If instead all symmetry-breaking terms were allowed,
with coefficients of order 1 in Planck units, the potential
would be given by the generic power-series expansion mentioned earlier.
While the shift symmetry justifies the flatness of the potential, it
does not suggest any particular form for it because we are in the large-field
regime.

Returning to the small-field regime, field theory
 with non-renormalizable terms
essentially ignored allows  only a few different types of term
for the variation of $V$. With the reasonable  assumption that 
one such term dominates over the relevant range of $\phi$,
and with the restriction $n<1$,
we arrive at essentially the models displayed in Table 2.
Details of these models, with extensive  references and possible
complications, are given
in \cite{treview}. One of these complications is the 
possibility, considered by several authors,
 that two terms need to be kept over the relevant range of $\phi$.
While this can happen, it is clear that the dominance of one term is the
generic situation in the sense that it will hold 
over most of the potential's parameter space.

 When the COBE normalization is imposed on the
prediction, the small-field requirement 
can generally be  satisfied  with physically reasonable values  of the
parameters. The only significant exceptions are  the
logarithmic potential with $c\sim 1$ (as in $D$-term inflation),
and the quadratic potential in the last line (more below on the latter),
which both require $\phi\sim\mpl$.

Given the restriction on $\phi$, the flatness conditions 
require that $V_0$ dominates the potential in all of the models,
 leading to simple expressions for $\epsilon$ and $\eta$.
The  contribution of
 gravitational waves is negligibly small in all of them,
 and 
 the formula for  $n$ is well approximated by
\be
n-1=2\eta
\label{nofvapprox}
\,.
\ee
There are models giving $n-1$ both positive and negative, but 
in the former case an observational value for $n$ does not
tell us much. For  $n<1$, in contrast, 
$n$ is a good discriminator between models. Some predictions are listed in
Table \ref{table2}.
 Except in the last row, the prediction
  depends on $N$ and is therefore scale-dependent. However, since
 $n$ is constrained to be close to 1,
the   scale-dependence is negligible
 over the cosmological
range $\Delta N\sim 4$, and accordingly one may set
$N=N\sub{COBE}$.
In the `new' inflation models, \eq{nresult}
gives
 a non-trivial lower bound on $N$, which would almost
exclude the $p=3$ model if this  1-$\sigma$ bound were taken seriously.

Another case of interest is the potential $V=V_0 -\frac12 m^2\phi^2
+\cdots$. 
More or less independently of the additional terms which stabilize
the potential, the vev of $\phi$ is  $\langle\phi\rangle\sim\sqrt{2V_0/m^2}
=[2/(1-n)]^{1/2} \mpl$.
Depending on the nature of $\phi$, this kind of inflation has been termed 
`natural', 	`topological' and `modular' (see for instance
\cite{bd} for a recent espousal of modular inflation).
In all cases 
the model is regarded as  implausible if  $\langle\phi\rangle$ is much
bigger than $\mpl$, which means that it is viable only if 
$n$ is not too close to 1.
 Our
 2-$\sigma$ bound
$n\gsim 0.9$ implies $\langle\phi\rangle \gsim 4.5\mpl$, which 
may perhaps be regarded as already disfavoring these models.

\begin{table}[t]
\caption{Some  field-theory models of inflation predicting a spectral
index $n<1$.}
\begin{center}
\footnotesize
\begin{tabular}{|lllll|}
\hline
model & potential & \multicolumn{3}{c|}{spectral index $n$} \\
&&& \multicolumn{2}{c|}{value of $N$} \\
& && $50$ & $20$ \\
\hline 
 &$ V_0(1+c\ln\phi)$ &  $1-\frac1N$ & 0.98 & 0.95 \\
'mutated' & $V_0(1-c\phi^{-2})$ &  $1-\frac3{2N}$ & $0.97$ & $0.93$ \\
'new' & $V_0(1-c\phi^4)$ &  $1-\frac3N$ & $0.94$ & $0.85$ \\
'new' & $V_0(1-c\phi^3)$ &  $1-\frac4N$ & $0.92$ & $0.80$ \\
'modular\dots' & $V_0-\frac12m^2\phi^2$ & 
{ $1-\frac{m^2\mpl^2}{V_0}$} && \\
\hline
\end{tabular}
\label{table2}
\end{center}
\end{table}

\section{Running-mass models of inflation}

So far we focussed on models giving a practically scale-independent 
spectral index. This seems to be a generic prediction of
inflation models based on spontaneously broken (global) supersymmetry.
As Stewart pointed out some years ago, the opposite is the case for
models based on softly broken supersymmetry \cite{running}.
In such models, the inflaton mass runs with scale, and  in the 
 linear log approximation the potential is
\be
V=V_0 -\frac12\frac{V_0}{\mpl^2} c \( \ln\frac{\phi}{\phi_*}-\frac12\)
\phi^2
\nonumber
\ee
This leads to 
\bea
\frac{n(k)-1}{2} &=& s e^{c\Delta N(k)} -c 
\nonumber\\
\Delta N& \equiv & \ln(k/k\sub{COBE} )
\eea
If $c$ is a gauge coupling, its 
 expected magnitude 
is

\be
|c|\sim 10^{-2} \mbox{\rm\  to\ } 10^{-1}
\nonumber
\ee
With $c$ in the upper part of this range, the spectral index can change
very significantly over the range $\Delta N\sim 4$ or so which corresponds
to cosmological scales.
(The other parameter $s$ controls end of inflation, and to avoid
severe  fine-tuning it should satisfy
$\phi\sub{end}\simeq\phi_*$.)

 The fit mentioned earlier \cite{cl00}
determines the region of $c$ and $s$ allowed by observation.
A gauge coupling $c\sim 0.1$ for the  inflaton  is allowed, giving 
potentially observable scale-dependence of the spectral index.

 \section{Gravitino creation from the vacuum}

There are strong cosmological constraints on the 
 abundance of the gravitino, over most of the expected mass range.
The light, practically stable gravitino typically  predicted by 
gauge-mediated models of supersymmetry breaking must not overclose
the Universe, while the heavier gravitino of gravity-mediated models
must not interfere with nucleosynthesis. Only the very heavy gravitino
predicted by anomaly-mediated supersymmetry breaking  seems to be
free from cosmological constraints. 

It has long been known that  gravitinos are efficiently produced
by thermal collisions after reheating, and that to make 
 these thermal gravitinos cosmologically safe usually requires
  a low reheat temperature
and/or sufficient late-time entropy production. More recently, it has
been noticed \cite{kklv,gtr,latetime}
that gravitinos  may be produced  even more efficiently
 between the end
of inflation and reheating, through the oscillation of the  field that was
responsible for the inflationary
energy density. These gravitinos are created from the vacuum, through the
amplification of the vacuum fluctuation.

The  abundance of gravitinos created from the vacuum
 is determined
by the evolution equation of the 
relevant mode function (the function multiplying the creation operator).
There 
are separate mode functions
for helicity $1/2$ and $3/2$, as seen by a comoving observer in the expanding 
Universe. The evolution of the helicity $3/2$ mode function 
is essentially the same as for 
a spin $1/2$ particle,  whose effective mass is the field-dependent 
gravitino mass 
$\mgrav$ appearing in the Lagrangian \cite{mm}. 
(We shall denote its vacuum value by simply $\mgravvac$.)
This means
 \cite{lrs}
that the creation of helicity $3/2$ gravitinos from the vacuum takes place
 just after inflation,
with  number density $n\sim 10^{-2}\mgravcu$ just after creation.
Barring an unforeseen cancellation, the supergravity expression for the
potential requires that in the early Universe $|\mgrav|\lsim H$.
As a result creation of helicity $3/2$ gravitinos from the vacuum
is insignificant compared with 
 gravitino 
 production from thermal collisions.

 The evolution of the
helicity $1/2$ mode function is far more complicated. It was given first 
\cite{kklv,gtr} 
in the case that only a single chiral superfield is relevant.
This    one-field
case  was at first
investigated 
 \cite{kklv,gtr}
 in the approximation of unbroken supersymmetry in the vacuum.
Under the reasonable approximation  that the  oscillating
field has a quadratic potential derived from global supersymmetry,
it was found  that the creation of  the helicity $1/2$ gravitino 
is the same as the creation of the inflatino. This may be traced to
the fact that in this approximation there is gravitino-goldstino
equivalence, the inflatino being the 
goldstino of spontaneously broken global supersymmetry.
The  number density is now $n\sim 10^{-2} M^3$
where $M$ is the mass of the oscillating field. This is bigger than
the abundance 
of  helicity $3/2$ gravitinos, since $M>H$ is required for the field
to oscillate.

The problem with the   approximation of unbroken supersymmetry in the vacuum is
that it makes the  gravitino massless in the vacuum, leaving the
inflatino as a physical particle.
 For this reason, the one-field model was next
studied  \cite{latetime} under the assumption that the oscillating field also
breaks supersymmetry in the vacuum, making it presumably a 
modulus of string theory.
The  result is now very   different; 
unless reheating intervenes,
gravitino
creation continues
 until the `intermediate' epoch, defined as the epoch at which 
 the energy density is of order 
\be
M\sub S^4\equiv 
3\mpl^2 \mgravvac^2
\ee
corresponding to Hubble parameter $H=\mgravvac$.
 (It was  assumed that
 inflation ends before the intermediate epoch,
as is the case in a wide class of inflation models.)

The number density  of gravitinos, just after creation
ends, is again 
$n\sim 10^{-2} M^3$ (we discount for the moment the case that 
the energy density of created gravitinos becomes significant). 
If $M$ is bigger than $\mgravvac$, 
 this
 {\em late-time gravitino creation} is   more efficient
than creation just after the end of inflation. 
However, because the oscillating field is now required to break supersymmetry
in the vacuum, $M$ cannot be many orders of magnitude bigger than
$\mgravvac$. As a result, it turns out that even 
 late-time gravitino creation cannot be as efficient as
thermal gravitino production. This one-field case is therefore of only
academic interest.

Recently, a formalism has been given
 \cite{kklv2} which describes the
 helicity $1/2$ gravitino in the presence of 
any number of chiral and gauge supermultiplets.
Using this formalism, it has recently been confirmed \cite{00grav}
 that late-time gravitino creation occurs in the generic
case, leading to a gravitino abundance at least as big as the one found
in the one-field model. I briefly explain how this comes about.

The  equation for the evolution of the helicity $1/2$
mode function $\theta$ is 
\bea
0 &=& \( \partial_0\partial_0 + k^2 + \hat B^\dagger \hat B + 2B_1\partial_0
 + \hat B' -ik \gamma_3\gamma_0 \hat A'
\)\theta\nonumber \\
&+& \( 2 B_1  - a\mgrav \) 
\( \partial_0 +\hat B -ik\gamma_3 \gamma_0 \hat A \) \theta\nonumber\\
&-&\frac{4ak^2}{\alpha} \Xi
\label{63a}
\eea
where a prime and $\partial_0$ both
denote differentiation with respect to conformal
time, $\diff/\diff x_0\equiv a\diff/\diff t$.
 In this equation, 
 $\hat A=A_1+\gamma_0 A_2$ and $\hat B
=B_1 + \gamma_0 B_2$, where 
\bea
A_1 &\equiv & \frac{p-3\mpl^2\mgravsq}{\rho+
3\mpl^2\mgravsq} \label{a1first}\\
A_2 &\equiv&  \frac{2\mpl^2\dot \mgrav}{\rho+
3\mpl^2\mgravsq} \label{a2first}\\
B_1 &\equiv& \frac{3a}{2} \( -H A_1 + \mgrav A_2\) \label{a3}\\
B_2 &\equiv& -\frac a 2 \[ 3H A_2 + \(1+ 3 A_1\) \mgrav\] \label{a4}
\eea
An  over-dot denotes $\diff/\diff t$, $\rho$ is the energy density, and
$p$ is the pressure.
 The energy density and pressure appear because they determine
 the Einstein tensor of  the Universe, which appears
because we are evolving the gravitino mode function in curved spacetime.
At least in the cases studied so far, the oscillation of $p$ caused by the
oscillating field is the dominant cause of gravitino creation.

The last  term involves
\be
\alpha = \rho+ 3\mpl^2\mgravsq
\ee
and $\Xi$ which is a linear combination of fermion fields, orthogonal
to the combination that is eaten by the gravitino to acquire mass.
The coefficient of each fermion field is a function  of the scalar fields
to which it couples.

The equation holds for each momentum $k/a$, where $a$ is the scale
factor of the Universe. Flat spacetime field theory holds during an
 `initial' era and a `final' era. Up to  slowly-varying pre-factors,
the initial condition is $\theta=
\exp(-ikx_0)$, corresponding to the vacuum, and the final 
occupation number  $|\beta|^2$ is read off from (while the gravitino
remains relativistic) from $\theta
=\alpha \exp(ikx_0) + \beta\exp(-ikx_0)$.
There is negligible creation in some  adiabatic regime
$k\gsim k\sub{max}$, in which 
$\theta= \exp(-ikx_0)$ holds at all times.

If the only relevant fields form a  chiral supermultiplet,
 $\Xi$ vanishes. This is the one-field case mentioned earlier,
and it gives  $k\sub{max}\sim a\sub{int}M$,
 where $M$ is the mass of the oscillating
field and the subscript denotes the intermediate epoch.
If the relevant fields 
form two chiral supermultiplets, $\Xi$ can be expressed in 
terms of  $\theta$; this case was considered in detail in \cite{00grav},
and shown to lead to the {\em same} estimate for $k\sub{max}$.
 In the general case,
$\Xi$ is an independent quantity, and
the evolution of $\theta$ and the spin $1/2$ fields is given
by a system of 
coupled equations. Without solving the system,
the fact that $\Xi$ is an
independent quantity means that, barring accidental cancellations,
 a lower 
bound on the gravitino abundance will be obtained by setting $\Xi=0$.
This leads \cite{00grav} to the estimate
\be
k\sub{max} \sim  a\sub{crea} M
\ee
where the subscript denotes the epoch when gravitino creation ends;
it is the intermediate epoch or the epoch of reheating, whichever is
earlier.

Because they are fermions, the 
abundance of created gravitinos can be estimated in terms of 
 $k\sub{max}$, and the occupation number $|\beta|^2$ for that wavenumber,
\bea
n &\sim & \frac1{2\pi^2} \int^{k\sub{max}}_0 a^{-3} |\beta_k|^2
 \diff k \nonumber \\
&\sim & 10^{-2} |\beta|^2(k\sub{max}/a)^3 \label{11}
\eea
The corresponding energy density is $\rho\sim (k\sub{max}/a) n$,
and it cannot be bigger than the total. This leads to the estimate
\cite{00grav}
\be
n \sim \min \{ 10^{-2} (k\sub{max}/a\sub{crea})^3,
\rho\sub{crea}(a\sub{crea}/k\sub{max}) \}  (a\sub{crea}/a)^3,
\ee

In the case of gravity-mediated supersymmetry breaking, 
 nucleosynthesis requires \cite{ns}
\be
n/s\lsim 10^{-13}
\label{nconst}
\ee
where $s$ is the entropy density.
Thermally produced gravitinos are subject to this constraint, 
and keeping only them it  requires \cite{ns}
\be
\gamma T\sub R\lsim 10^9\GeV
\label{thermcon}
\ee
Here,  $T\sub R$ is the reheat temperature, defined as the 
temperature just after all 
or most of the energy in the oscillation is converted into
radiation, and  $\gamma^{-1}\geq 1$ is the entropy increase, if any, 
after  reheating.

Now consider instead
the gravitinos created from the vacuum.
We can work out $n/s$ at nucleosynthesis,
remembering that  $n\propto a^{-3}$ is proportional to
$\rho$ until reheating, and to
 $s\gamma$  thereafter.
Considering only the non-relativistic regime, we
find that gravitinos created from the vacuum lead to the
following nucleosynthesis constraint if $T\sub R\lsim M\sub S$
\be
\gamma T\sub R\lsim \max \{10^{-11}M\sub S^4/M^3,10^{-13} M\}
\label{con1}
\ee
If instead $M\sub S\lsim T\sub R\lsim M$,
we find again the second of the previous constraints,
\be
\gamma T\sub R\lsim 10^{-13} M \label{con2}
\ee 
Finally, if $T\sub R\gsim M\sub S$ and $T\sub R\gsim M$ we find
\be
T\sub R \gsim 10^4\gamma^{1/3} M \label{con3}
\ee

The requirements that gravitinos do not spoil nucleosynthesis are
\eq{thermcon}, plus the appropriate one of \eqss{con1}{con2}{con3}. 
Before considering these constraints, we have to consider
the possible entropy increase $\gamma^{-1}$. The most efficient mechanism
of entropy release is  thermal
inflation  \cite{thermal} 
(or some other type of inflation occuring after the intermediate epoch).
One bout of thermal inflation  gives huge entropy release, roughly
$\gamma^{-1}\sim 10^{15}$, and there could be more than one.
If there is no thermal inflation,
significant entropy release can come only from an era of matter domination by
an unstable particle. (For simplicity we exclude the case of two or more
such eras.) Then 
\be
\gamma^{-1}\sim T\sub{eq}/ T\sub{decay}
\label{gammaeq}
\ee
where $T\sub{decay}$ is the  temperature  just after the particle
decay (ie., the final reheat temperature) and $T\sub{eq}$ is the temperature
 just before the era of matter domination.
In this case, $\gamma T\sub R \sim T\sub{decay} (T\sub R/T\sub{eq} )$.
The era of matter domination must end before nucleosynthesis, corresponding to
$ T\sub{decay} \gsim 10\MeV$,
which implies 
\be
 \gamma T\sub R \gsim 10\MeV
\,.
\label{con4}
\ee

The region forbidden  by nucleosynthesis,
in the space of the three parameters $\gamma$, $T\sub R$ and $M$,
is given by \eq{thermcon}, the relevant one of \eqss{con1}{con2}{con3},
plus \eq{con4} if there is no thermal inflation.
When $T\sub R\lsim M\sub S$, the boundary of this region depends
only on $\gamma T\sub R$ and $M$.
For higher values of $T\sub R$ it depends separately
on $\gamma$ and $T\sub R$.
If  $\gamma\lsim 10^{-12}$ there is no forbidden region.
 (In particular, there is no forbidden region if thermal
inflation occurs.) If
  $M\lsim 10^7\GeV$, the forbidden region is the same as for thermally
produced gravitinos. Otherwise, gravitinos created from the vacuum
rule out  a significant 
 portion of the parameter space, beyond what is ruled out by
thermally produced gravitinos.

\section{Where are we going with inflation models?}

By building and testing models of the early Universe, we obtain
 a unique window on the nature of the
fundamental interactions. This is especially true of inflation 
model-building, because the crucially important curvature perturbation,
once generated, is frozen in until well after nucleosynthesis.

In a few years   we shall 
know $n(k)$   with accuracy $\pm 0.01$.  This is 
the  only observable {\em function} relating to physics far beyond
the Standard Model!
The  measurement of $n(k)$, plus other constraints like gravitino abundance,
 will rule out most of the presently 
 existing inflation models. Depending on whether or not $n$ is significantly
different from 1, and on how much our understanding of string-derived field
theory progresses, one
 model  may have been selected as the best candidate.

The {\em  next frontier} will be to discover how the inflaton
sector talks to the Standard Model sector. There must indeed be 
communication 
because the inflaton field must decay into SM radiation ('reheating').
As a result, given continued progress (which might be the rub), top-down
inflation model-builders  {\em will} eventually meet up with
bottom-up extenders of the Standard Model!

\end{document}